\documentstyle[prl,aps,epsfig]{revtex} 

\begin{document}

\draft                    

\title{
         Ab initio results for the electronic structure of
         \textbf{$C_{50}Cl_{10}$}.}

\author{
         Steven W.D.~Bailey$^{1}$
         and
         Colin J.~Lambert$^{1}$
        }

\address{$^{1}$ Department of Physics, Lancaster University,
         Lancaster, LA1 4YB, UK}


\maketitle


\begin{abstract}
In this paper we use ab initio density functional theory (DFT) to
calculate the electronic properties of $C_{50}Cl_{10}$. In
comparison with the unstable $C_{50}$ which has a small
$t_{lu}(LUMO)$ - $h_{u}(HOMO)$ energy gap and a high total free
energy compared with $C_{60}$, the belt of chlorines atoms
stabilize the $C_{50}Cl_{10}$ fullerene by increasing the energy
gap to approximately that of $C_{60}$ and lowering the total free
energy. We also examine the effects of inter-cage separation on
the band structure for infinite periodic $C_{50}Cl_{10}$ chains
where a high degree of dispersion is found to persist for
separations beyond the predicted $C_{60} - C_{60}$ distance of
closest approach.
\end{abstract}
\begin{figure}
         \epsfxsize=0.6\columnwidth
         \epsffile{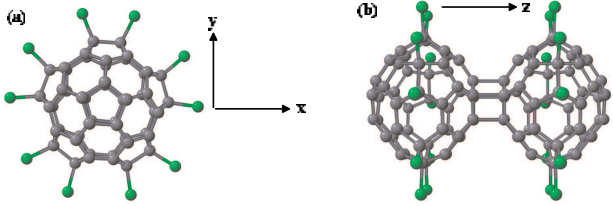}
         \vspace{0.05\columnwidth}
\caption{Relaxed atomic coordinates of $C_{50}Cl_{10}$. The
chlorine atoms form a fan around the equator of the $C_{50}$
molecule (\textbf {a}). The proposed chain of $C_{50}Cl_{10}$ in
the z direction (\textbf {b})}
 \label{Fig1}
\end{figure}

A new example of the fullerene family, $C_{50}Cl_{10}$ has
recently been produced as a stable solid \cite{Xu2004}. The
fullerenes smaller than the $C_{60}$ buckyball are very difficult
to produce in a stable form as the structure breaks the isolated
pentagon rule (IPR) \cite{Kadish2002}. In  $C_{50}$ there are five
sets of adjacent pentagons and this together with the high
curvature of the molecular surface means it is highly unstable
\cite{Guo1992,Kroto1987,Heath1998}. Xie et al have overcome this
instability by attaching chlorine atoms around the $C_{50}$
equator as seen in Fig.~\ref{Fig1}(a). In what follows, we examine
the electronic properties and stability of $C_{50}Cl_{10}$, along
with the electronic properties of chains of $C_{50}Cl_{10}$
molecules. To model $C_{50}Cl_{10}$ we use a double zeta basis set
with a cutoff radius of $150$ Ry to describe the carbon and
Chlorine atoms within the SIESTA \cite{siesta97} implementation of
DFT, where nonlocal norm-conserving pseudopotentials describe the
core electrons and a linear combination of atomic orbitals the
valence electrons. The relaxed, isolated molecule is shown in
Fig.~\ref{Fig1}(a). In the case of a chain we concider molecules
aligned along the z axis as shown in Fig.~\ref{Fig1}(b).\\%
\textbf{TABLE 1.} compares the Fermi energy $E_{F}$(eV), the total
free energy $E_{tot}$(eV) and the $t_{lu}(LUMO)$ - $h_{u}(HOMO)$
energy gap $E_{\Delta}$(eV) between the $C_{60}$, $C_{50}$
molecules and $C_{50}Cl_{10}$ as a linear chain with varying
inter-cage separations in ($\dot{A}$), up to the isolated molecule.\\%
\textbf{TABLE 1.}\\%
\begin{tabular}{ |c |c |c |c | c | c| c| }
  \hline
  Case & a & b & c & d & e & f\\
  &$C_{60}$ mol& $C_{50}$ mol&$C_{50}Cl_{10} 2.5\dot{A}$&$C_{50}Cl_{10} 2.9\dot{A}$&
  $C_{50}Cl_{10} 3.4\dot{A}$&$C_{50}Cl_{10}$ mol\\
  \hline
  $E_{F}$(eV)& -4.61 & -7.04 & -5.41 & -5.41 & -5.47 & -5.67\\
  $E_{tot}$(eV)& -9201 & -7670 & -11751 & -11751 & -11751 & -11750\\
  $E_{\Delta}$(eV) & 1.80 & 0.16 & --- & --- & ---
   & 1.80\\
  \hline
\end{tabular}\\%

\begin{figure}
         \epsfxsize=0.75\columnwidth
         \epsffile{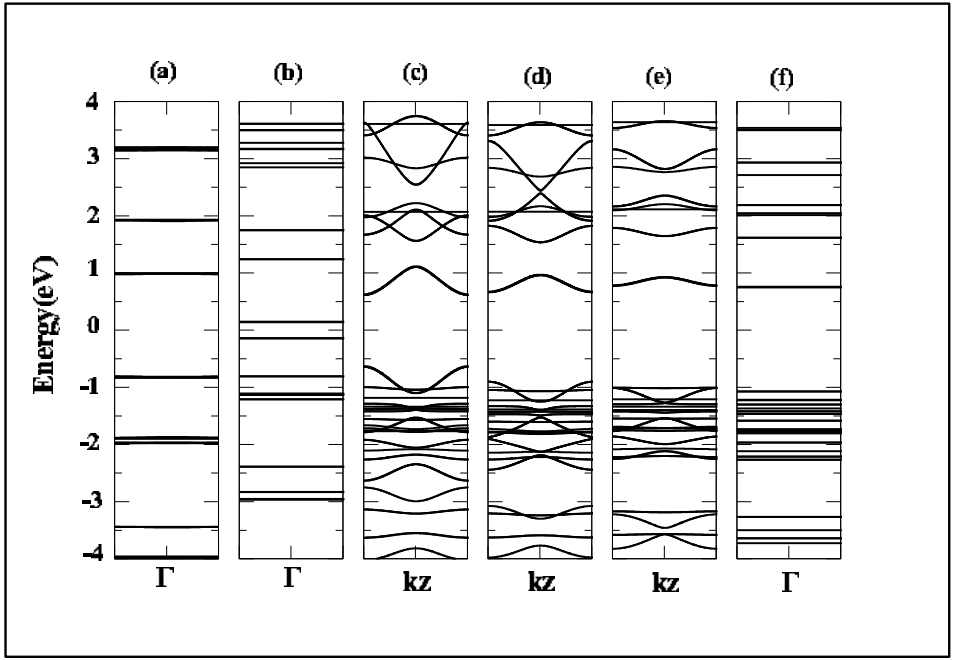}
        \vspace{0.05\columnwidth}
\caption{For the structures (a - f) defined in Table 1, the energy
levels of the isolated molecules are seen in cases (a,b and f) and
the electronic band structures for wavevectors $kz$ in the z
direction as defined in Fig.~\ref{Fig1}, in cases (c - e).}
 \label{Fig2}
\end{figure}

The table shows that the unstable $C_{50}$ has a high $E_{tot}$ of
$-7670$eV in comparison to that of $-9201$eV for the $C_{60}$
molecule but that the belt of ten $Cl$ atoms stabilizes the
$C_{50}Cl_{10}$ to produce a lower $E_{tot}$ of $-11751$eV. This
is reflected in the result for the $t_{lu}(LUMO)$ - $h_{u}(HOMO)$
energy gap, ($E_{\Delta}$), where the energy gap of $0.16$eV for
$C_{50}$ opens to $1.80$eV, matching that of $C_{60}$, for the
$C_{50}Cl_{10}$ molecule. It is noted that the value of
$E_{\Delta} = 1.80$eV for $C_{60}$ is in very good agreement with
the experimental values of $E_{\Delta}$ between ($
1.6 - 1.8$)eV\cite{D&D1996}\\%
Fig.~\ref{Fig2} shows the energy levels of the isolated molecules
in cases (a,b and f) and the electronic band structures of
$C_{50}Cl_{10}$ chains with the different inter-molecule
separations in cases (c - e), of Table 1. In each case the energy
is measured from the $E_{F}$ for each relaxed structure studied.
In cases(c - e) the band structure is shown for a wavevector $kz$
in the z direction as defined in Fig.~\ref{Fig1}. The $C_{60} -
C_{60}$ distance of closest approach is predicted to be
approximately $3.0\dot{A}$ \cite{Rochefort03} therefore the range
of cage separations for cases (c - e) were selected to cover this
interval and beyond. The band structures for the infinite periodic
chains of $C_{50}Cl_{10}$ for cases (c - e) show the dispersion of
the $t_{lu}(LUMO)$ and $h_{u}(HOMO)$ bands persists up to and
beyond a $3.4\dot{A}$ separation and that the $C_{50}Cl_{10}$
chain
is an insulator.\\%
In conclusion we have used DFT calculations to investigate the
electronic properties of the newly produced, solid $C_{50}Cl_{10}$
fullerene. We find that the belt of chlorine atoms around the
equator of the $C_{50}$ stabilize the structure by dramatically
lowering the total free energy of $C_{50}Cl_{10}$ and increasing
the $h_{u}(HOMO)$ - $t_{lu}(LUMO)$ gap to match that of $C_{60}$.
The $t_{lu}(LUMO)$ and $h_{u}(HOMO)$ bands show a marked
dispersion even for separations of the $C_{50}Cl_{10}$ cages in
excess of the predicted closest $C_{60} - C_{60}$ approach of
$3.0\dot{A}$. This points to some interesting properties arising
from carbon nanotube peapods such as $(12,12)@C_{50}Cl_{10}$.


 \end{document}